# Enhancement of Photoemission on P-type GaAs using Surface Acoustic Waves

Boqun Dong,[1,*] Andrei Afanasev,[2] Rolland P. Johnson,[3] Mona E. Zaghloul[1]

[1] School of Engineering and Applied Science, [2] Department of Physics, The George Washington University, Washington, DC 20052, USA. [3] MuPlus, Inc., Newport News, VA 23606, USA.
*Email: dongbq@gwmail.gwu.edu

**Abstract.** We demonstrate that photoemission properties of GaAs photocathodes (PCs) can be altered by surface acoustic waves (SAWs) generated on the PC surface due to dynamical piezoelectric fields of SAWs. Simulations with COMSOL indicate that electron effective lifetime in p-doped GaAs may increase by a factor of 10x to 20x. It implies a significant, by a factor of 2x to 3x, increase of quantum efficiency (QE) for GaAs PCs. Essential steps in device fabrication are demonstrated, including deposition of an additional layer of ZnO for piezoelectric effect enhancement, measurements of I-V characteristic of the SAW device, and ability to survive high-temperature annealing.

## 1. Introduction

Photocathodes (PCs) are used as a source of electrons for numerous applications that include photomultipliers, electron microscopes and particles accelerators[1-3]. If spin-polarized electrons are required for applications, a standard choice is III-V semiconductors activated to negative electron affinity (NEA)[4]. The three-step photoemission mechanism[5] of p-type GaAs PC with NEA coating is shown in Figure 1(a). The photovoltaic effect and semiconductor properties of GaAs are utilized to achieve photoexcitation (step I) and electron transportation (step II). The NEA coating is used to lower the vacuum level to below the conduction band minimum for electron emission (step III)[5]. High quantum efficiency (QE) of PCs is an important requirement, for example, for an Electron-Ion Collider for next-generation fundamental research in nuclear physics[6], where spin-polarized electron beams are needed to achieve research objectives. Recent progress[7] for increasing the QE of a spin-polarized GaAs/GaAsP superlattice electron source was due to the use of a distributed Bragg reflector that was designed to increase the photon absorption in a thin active layer of the PC.

In this work we study the effect of piezoelectric fields on the minority carriers in p-doped GaAs PCs. This research was motivated by findings[8] that piezoelectric fields generated in GaAs by surface acoustic waves (SAWs) suppress recombination and lead to extended electron lifetimes in this system. This is achieved by using SAWs to periodically bend the energy bands and to spatially separate electrons and holes[8-10] as shown in Figure 1(b). Two of us[11] suggested earlier that suppression of charge-carrier recombination due to SAW may improve PC performance.

In this present work, we use COMSOL Multiphysics as a tool to build models and to simulate step I and step II of the photoemission mechanism. The QE of the emission process in step III is derived based on theoretical principles. All the work is completed for both a bulk GaAs structure and a thin film GaAs structure.

## 2. Results and Discussion

2.1 Bulk p-type GaAs

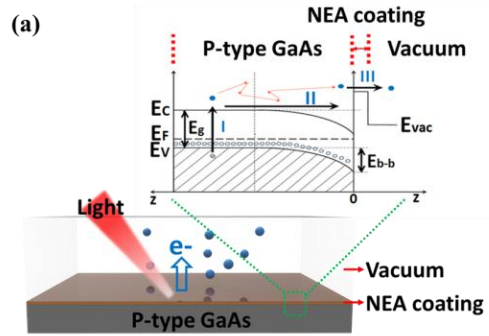

Fig. 1. (a) The bottom is a three-dimensional (3D) schematic view showing the photoemission from p-type doped GaAs surface with a thin NEA coating. The inset is a band diagram demonstration of the three-step photoemission mechanism: I. photoexcitation; II. transport; III. emission from the surface.

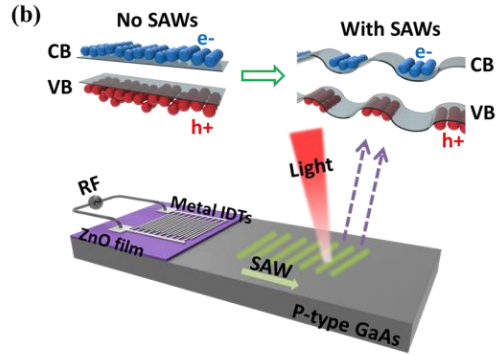

Fig 1. (b) The 3D schematic view at bottom shows the concept and structure of our IDTs device used to generate SAWs on p-type doped GaAs substrate. Top left - the result of photoexcitation without SAWs. Top right - the band bending effect caused by SAWs. In this case, the electrons and holes are spatially separated, thus the recombination is suppressed.

Figure 2(a) shows the cross-section view of the simulation structure, which is the same as the 3D model shown in Figure 1(b). The substrate is highly doped p-type GaAs with doping concentration of $5 \times 10^{18}$ cm$^{-3}$. A ZnO film is placed on top of the GaAs to enhance the generation of SAWs due to its strong piezoelectric effect. Two pairs of interdigital transducers (IDTs) made of aluminum are deposited on top of the ZnO layer on each side of the illumination area and used to generate SAWs. IDTs no.1 and no.3 are grounded. A radio frequency AC voltage is applied to IDTs no.2 and no.4 that is expressed by the following equation:

$$V_{in} = V_0 \times sin(2\pi \times f_0 \times t), \quad (1)$$

where $V_0$ is chosen as 1.0 V, and $f_0$ is set to 300 MHz during simulation. The SAW is treated as a Rayleigh Wave in the simulation. The wave propagating speed is set to 2760 m/s. Thus the wavelength of SAW is calculated to be 9.2 μm. The thickness of substrate is 18.5 μm and the width is 190 μm. The IDTs are each 2.3 μm wide × 2.3 μm high. The distance between each IDT is also 2.3 μm.

In simulation, light illumination is applied to the center of the top of the bulk GaAs, as indicated in the purple rectangular area in Figure 2(a). The width of the area is 50 μm and the depth is 2 μm. Photoexcited electrons are created under illumination by a laser beam. The power of the laser is set to 5 W/mm$^2$. The wavelength of incident photons is set to 600 nm, thus the photon energy is 2.07 eV.

The results shown in Figure 2(b) and Figure 2(c) are calculated and extracted along the red cutline indicated in Figure 2(a). The cutline on the top surface of the highly p-doped GaAs is in the horizontal direction, which is also the propagation direction of the SAW. Results show that recombination rates are reduced by $10^2$ - $10^3$ times and electron concentrations increased about 14 times at the surface of the photoemission area when SAWs propagate along the surface of the GaAs.

In order to verify the dependence of the enhancement of the electron concentration on SAW intensity, the amplitude of the applied AC voltage is swept from 0.2 V to 2.0 V in steps of 0.1 V. Simulation results are shown in Figure 2(d). With the increase of amplitude of AC voltage and thus the SAW intensity, more photoexcited electrons are able to reach the top surface of the GaAs. The maximum applied voltage $V_0$ in simulation was 2.0 V. A description of this voltage limit is given in Section 3.

Next, to investigate the effect of SAWs under different wavelengths of incident photons, another two sets of simulations were conducted with results shown in Figure 2(d). According to the results, both higher energy photons (600 nm, 2.07 eV) and lower energy photons (800 nm, 1.55 eV) are able to generate more electrons with the help of SAWs. For all three wavelengths of photons, similar enhancements are achieved under SAWs. Among them, photons with 700 nm wavelength have the highest photo-generated electron concentration, which is consistent with the photo-absorption property of GaAs.

Results in Figure 2(e) are calculated and extracted along the green cutline that is the photon absorption depth in GaAs in vertical direction, as indicated in Figure 2(a). In Figure 2(e), the leftmost side (0 nm) of the x-axis refers to the top surface of GaAs, and the rightmost side (2000 nm) refers to the bottom of absorption area. Results show a stable and significant enhancement from the surface to the bottom of the absorption area since the wavelength of the SAW (9.2 μm) is bigger than the depth of the absorption area (2 μm) in this simulation.

In order to investigate the dependence of electron concentration enhancement on the SAW wavelength, another simulation structure was modeled and calculated. In this model, the size of IDTs is greatly reduced to make the SAW wavelength (0.5 μm) smaller than the absorption depth (2 μm). This simulation result is shown in Figure 2(f). The blue line is same as the one shown in Figure 2(e), indicating the electron concentrations without SAWs. The trend of the red line, however, changed in this new simulation. The SAW enhancement is reduced after the absorption depth exceeds half the SAW wavelength.

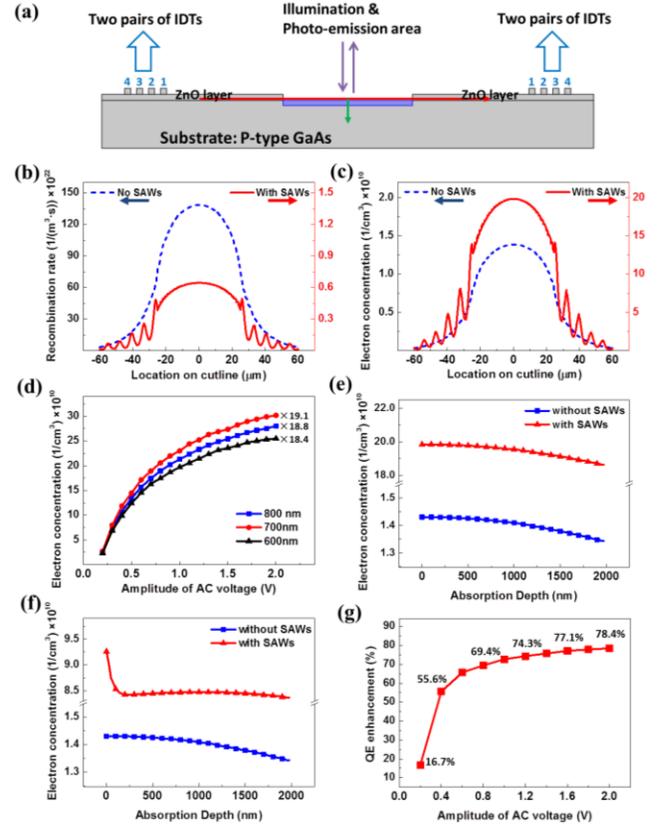

Fig. 2. (a) Simulation structure of bulk p-type doped GaAs with IDTs. (b) Comparison of recombination rates with and without SAWs ($V_0$ = 1.0 V). (c) Comparison of surface electron concentrations with and without SAWs ($V_0$ = 1.0 V). (d) Electron concentrations vs. AC voltage and photon wavelengths of incident light. (e) Electron concentrations along absorption depth for SAW wavelength ($\lambda_{SAW}$ = 9.2 μm) that is bigger than the absorption depth (2.0 μm) in bulk GaAs. (f) Same as (e) but for $\lambda_{SAW}$ = 0.5 μm, that is smaller than the absorption depth. (g) Enhancement of QE under different amplitudes of AC voltage.

According to the simulation results that recombination is suppressed and electron concentration increased, the lifetime $\tau$ of electrons is increased by the same multiple based on the following equation:

$$\frac{n_{SAW}}{n} = \frac{\tau_{SAW}}{\tau}, \quad (2)$$

that follows from the relation between the equilibrium concentration of excess carriers ($n$), carrier lifetime ($\tau$) and photoinjection rate ($r_{in}$): $n = \tau \cdot r_{in}$. Next, based on:

$$L = \sqrt{\tau \times D}, \quad (3)$$

the diffusion length $L$ extends with carrier lifetime $\tau$.

The QE can be described by the formula[12]:

$$QE = (1-R)\left(\frac{\alpha_{PE}}{\alpha}\right)\left(\frac{P_E}{1+\frac{l_\alpha}{L}}\right), \quad (4)$$

where $R$ is the reflectivity, $\alpha_{PE}/\alpha$ is the fraction of electrons excited above vacuum level, $l_\alpha$ is the absorption depth, $L$ is the diffusion length, and $P_E$ is the probability that electrons penetrate the NEA barrier and escape into the vacuum, after reaching the sample's surface. For this bulk GaAs model, the photoexcited electron is assumed to have only one chance to hit the surface for photoemission. Hence, the diffusion length $L$ is the only parameter in this formula that will change with the application of SAWs, and thus the SAW enhanced QE can be derived by the equation:

$$\frac{QE_{SAW}}{QE_0} = \frac{1+\frac{l_\alpha}{L_0}}{1+\frac{l_\alpha}{L_{SAW}}}, \quad (5)$$

For the calculation, $L_0$ is set to 1.5 μm according to literature[13]. $l_\alpha$ is 2.0 μm as used in simulation. Based on the increased electron concentrations shown in Figure 2(d), $L_{SAW}$ is calculated using equation 2 and equation 3. Then the QE enhancement ($QE_{SAW}/QE_0 - 1$) for varied amplitudes of AC voltage is calculated and shown in Figure 2(g). The effect in bulk GaAs is strongly pronounced for $l_\alpha \gg L$, i.e., for photon energies approaching the band-gap, leading to enhanced QE for the photons of longer wavelengths.

2.2 Thin film p-type GaAs

In this section, the structure of a thin film of highly doped p-type GaAs placed on top surface is modeled and simulated. As shown in Figure 3(a), the substrate is p-type GaAs with doping concentration of $1 \times 10^{18}$ cm$^{-3}$. A layer of GaAs$_{0.7}$P$_{0.3}$ (thickness: 2 μm) is placed on GaAs substrate to create a potential barrier between surface and substrate. Then a thin layer of p-type GaAs is placed on top of GaAs$_{0.7}$P$_{0.3}$. The thickness is 100 nm and the doping concentration is $5 \times 10^{18}$ cm$^{-3}$. The structure of IDTs, the parameters of SAWs, and the applied AC voltage are the same as those with bulk p-type GaAs model (Figure 2(a)).

In this simulation, illumination is applied to the thin film of p-doped GaAs. The power of the laser beam is 5 W/mm$^2$. The wavelength of incident photons is set to 800 nm, thus the photon energy is 1.55 eV. Corresponding simulation results are shown in Figure 3(b) to 3(d). As in the bulk GaAs simulation, the recombination is greatly suppressed by $10^3$ times because of the SAWs. The amount of electron concentration at the surface is lower compared with results shown in Figure 2(c) and 2(d), because the absorption area of thin film GaAs is smaller than that of bulk GaAs. However, the electron concentrations are still increased more than ten times due to SAWs.

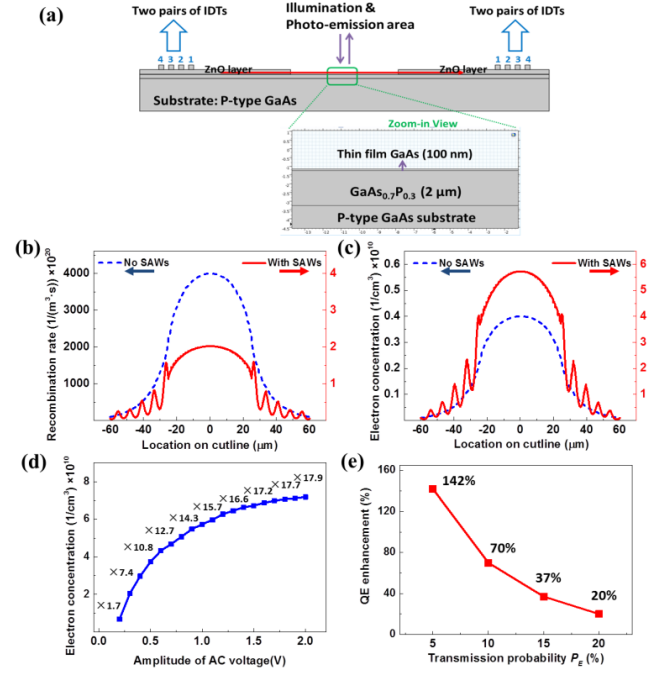

Fig. 3 (a) Simulation structure of thin film GaAs with IDTs. (b) Comparison of recombination rates in GaAs thin film with and without SAWs ($V_0 = 1.0$ V). (c) Comparison of surface electron concentrations with and without SAWs ($V_0 = 1.0$ V). (d) Increase of electron concentrations caused by SAWs for different amplitudes of AC voltage. (e) Enhancement of QE caused by SAWs under different transmission probability $P_E$.

Figure 3(d) shows the dependence of enhancement on SAW intensity. The x-axis is the amplitude of applied AC voltage swept from 0.2 V to 2.0 V. The y-axis shows the electron concentration at the surface of the p-doped GaAs thin film under the effect of SAWs. The numbers indicated above the curve refer to different enhancement of surface electron concentrations with varied SAW intensities.

For the thin film GaAs model, the photoexcited electron has multiple chances $N$ to hit the surface for photoemission because of two reasons: first, $d \ll L$; second, the potential barrier created by the GaAs$_{0.7}$P$_{0.3}$ layer helps push back electrons for another try. Therefore, according to equation 4, the SAW enhanced QE can be written in the following equations:

$$\frac{QE_{SAW}}{QE_0} = \frac{1-(1-P_E)^{N_{SAW}}}{1-(1-P_E)^{N_0}}, \quad (6)$$

$$N = \frac{L}{2 \times d} + 1, \quad (7)$$

For the calculation, $d$ is 0.1 μm as used in simulation, $L_0$ is 1.5 μm according to literature[13], and $L_{SAW}$ is 6.4 μm as calculated using equation 2 and equation 3 based on the result (at $V_0 = 2.0$ V) shown in Figure 3(d). By sweeping $P_E$ from 5% to 20%, the QE enhancement is calculated and plotted in Figure 3(e). Results show that more dramatic enhancement from SAWs is achieved for cases of smaller transmission probability $P_E$.

### 3. Fabrication and Experiment

Figure 4(a) and 4(b) present the outcome of the IDT device fabrication on a ZnO film and highly p-doped GaAs substrate, clearly showing the metal IDT fingers are well shaped. The metal surface is a little rough due to the quality of chamber vacuum during the e-beam evaporation process. The ZnO surface, however, is smooth and clean as needed for generating SAWs.

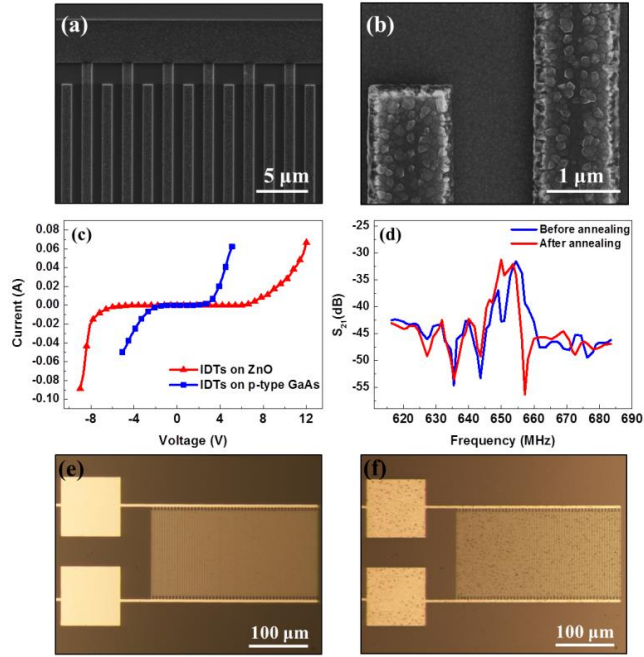

Fig. 4 (a) Scanning electron microscope (SEM) image of the top surface view of IDT fingers placed on the ZnO film. (b) High-resolution SEM image showing the surface morphology of the aluminum IDT finger and the c-axis oriented ZnO layer. (c) I-V characteristics of IDTs fabricated on different materials. (d) Comparison of transmission property $S_{21}$ of SAWs generated before and after annealing. Optical microscope images of the IDTs taken (e) before and (f) after annealing.

Figure 4(c) shows two measured I-V curves of IDT devices fabricated on p-type GaAs substrate with and without the ZnO film. Results show that the IDT directly deposited on the GaAs substrate has a safety limit of 2.5 volts. Above this limit, the current increases rapidly and the device is in danger of being damaged. That is the reason we set the sweeping range from 0 to 2 volts in simulations. Next, we can see the voltage limit of the device with the ZnO film increases from 2 volts to 6 volts. This is because ZnO has larger energy band-gap than GaAs, thus it is able to resist higher voltage and is more difficult to break down.

Additionally, since many photoemission applications such as PCs have heat processing during operation, we put our device into an annealing system (500 °C, 10 minutes) to test its robustness. By comparing Figure 4(e) and 4(f), it is clear that both the metal IDTs and the surface of ZnO film are not damaged during the annealing process. The only change is that some small dark spots appeared on the surface after annealing. Most of the dots are small dirt and defects that were originally located on surface and are exposed after annealing, thus will not reduce the performance of the SAW. As shown in Figure 4(d), the peak value and bandwidth of the transmission property $S_{21}$ are consistent before and after annealing, proving the SAW is still able to be generated and is transmitted successfully after the high-temperature annealing process.

### 4. Summary

In summary, we have demonstrated via simulations that SAWs applied to p-doped GaAs extend the minority-carrier (electron) lifetimes by over an order of magnitude. Importantly for the applications in spin-polarized electron sources, the extension of electron lifetimes is accompanied by extension of their polarization lifetimes, as seen in an experiment[14]. Further calculations predict that the QE of GaAs PCs would increase by a factor of 2 - 3 depending on the device geometry and NEA coating properties. In particular, reduction of the thin-film PC operational lifetime can be potentially mitigated by the SAW-enhanced electron lifetimes. Essential components of the prototype device were fabricated and showed high-quality SAW generation with an additional layer of ZnO applied under IDTs. We measured the device's I-V characteristic, and its performance after high-temperature annealing.

The described method to increase PC QE is expected to have a number of important applications: *High-brightness unpolarized electron sources* for high-energy physics, whereby increasing QE without changing the laser beam spot size on a PC surface provides higher beam brightness. *Polarized electron sources for particle accelerators[3]*, where the proposed method of modifying electron dynamics can be combined with a distributed Bragg reflector method to increase light absorption[7] for high-intensity low-emittance beams for fundamental nuclear physics research[6]. *Electron microscopy*, for which electron beam emittance can be reduced for the same beam intensity due to reduced laser-spot size. *Photomultipliers and photon detectors*, for which the spectral range can be improved, especially the sensitivity for the photon energies approaching the semiconductor's band gap.

**Acknowledgments**

Research partially supported by DOE HEP STTR grant DE-SC0017831. We appreciate the support from the nanofab facilities at National Institute of Standards and Technology (NIST) and The George Washington University (GWU).**References**

1. B. Lopez Paredes, H. M. Araujo, F. Froborg, N. Marangou, I. Olcina, T. J. Sumner, R. Taylor, A. Tomas, A. Vacheret, Astroparticle Physics 102, 56 (2018).
2. T.J. Chen, Y.J. Pei, Z.X. Tang, W.M. Li, Linear Accelerator Conf.(LINAC'18), Beijing, paper MOPO029, 90 (2018).
3. C. Hernandez-Garcia, P. G. O'Shea, and M. L. Stutzman, Physics Today 61(2), 44 (2008).
4. D.T. Pierce, F. Meier, Phys. Rev B13, 5484 (1976).
5. W.E. Spicer, Phys. Rev. 112, 114 (1958).
6. A. Accardi et al., Electron Ion Collider: The Next QCD Frontier - Understanding the glue that binds us all, Report number: BNL-98815-2012-JA; JLAB-PHY-12-1652, e-print arXiv:1212.1701, (2012).
7. W. Liu, Y. Chen, W. Lu, A. Moy, M. Poelker, M. Stutzman, and S. Zhang, Appl. Phys. Lett. 109, 252104 (2016).
8. C. Rocke, S. Zimmermann, A. Wixforth, J. P. Kotthaus, G. Böhm, and G. Weimann, Phys. Rev. Lett. 78, 4099 (1997).
9. J. B. Kinzel, D. Rudolph, M. Bichler, G. Abstreiter, J. J. Finley, G. Koblmuller, A. Wixforth, and H. J. Krenner, Nano letters 11(4), 1512 (2011).
10. B. Dong, S. Guo and M. Zaghloul, URSI Asia-Pacific Radio Science Conference (URSI AP-RASC), Seoul, 1921 (2016).
11. A. Afanasev and R.P. Johnson, Proc. Particle Accelerator Conference (IPAC 2011), New York, paper THP199 (2011).
12. W.E. Spicer, A. Herrera-Gomez, Photodetectors and Power Meters 2022, 18 (1993).
13. J. Nelson, *The Physics of Solar Cells,* World Scientific Publishing Company, 2003.
14. J. Wanner et al. Adv. Mater. Interfaces 1, 1400181 (2014).
5